\documentclass[twocolumn,showpacs,preprintnumbers,amsmath,amssymb]{revtex4}
\usepackage[dvips]{graphicx}
\usepackage{dcolumn}
\usepackage{bm}
\usepackage{color} 
\begin{document}

\title{Carrier Mobilities in $\delta $-doped Heterostructures}
\author{Y. Shao }
\author{S. A. Solin }%
\altaffiliation{Fax: 314-935-5983\\}
\email{solin@wustl.edu}
\affiliation{%
Washington University in St. Louis, Center of Materials Innovation, 
Department of Physics, St. Louis, Missouri 63130\\}%
\author{L R Ram-Mohan}
\affiliation{
Departments of Physics, Electrical and Computer Engineering, Worcester Polytechnic Institute, Worcester, MA 01609\\}%

\date{\today}

\begin{abstract}
For applications to sensor design, the product \textit{n$\times \mu {\rm }$} of the 
electron density $n $ and the mobility \textit{$\mu $} is a key parameter to be optimized for 
enhanced device sensitivity. We model the carrier mobility in a two 
dimensional electron gas (2DEG) layer developed in a $\delta $-doped 
heterostructure. The subband energy levels, electron wave functions, and the 
band-edge profile are obtained by numerically solving the Schr\"{o}dinger 
and Poisson equations self-consistently. The electron mobility is calculated 
by including contributions of scattering from ionized impurities, the 
background neutral impurities, the deformation potential acoustic phonons, 
and the polar optical phonons. We calculate the dependencies of \textit{n$\times \mu {\rm }$} on 
temperature, spacer layer thickness, doping density, and the quantum 
well thickness. The model is applied to $\delta $-doped quantum well 
heterostructures of AlInSb-InSb. At low temperature, mobilities as high as 
1.3x10$^{3 }$m$^{2}$/V$\cdot $s are calculated for large spacer layers (400 
{\AA}) and well widths (400 {\AA}). The corresponding room temperature 
mobility is 10 m$^{2}$/V$\cdot $s. The dependence of \textit{n$\times \mu {\rm }$} shows a maximum for a 
spacer thickness of 300 {\AA} for higher background impurity densities while 
it continues to increase monotonically for lower background impurity 
densities; this has implications for sensor design.

\end{abstract}

\maketitle

\section{\label{sec:level1}Introduction\protect}
In a number of magnetic sensor applications [1-2] the basic considerations 
for improved sensor performance are using ultra-thin films, of thicknesses 
$<$ 100 nm, and very high ($>$ 1 m$^{2}$/V$\cdot $s) room temperature 
mobility. However, it is also important to optimize the product of the 
carrier density, $n$, and the mobility, \textit{$\mu $}, in these devices since they have 
sensitivity that is proportional to the square of the mobility [3-4], and 
because high electron concentrations $n$ and electron mobility \textit{$\mu {\rm g}$}end to reduce 
the temperature variation of the sensor output [1, 4]. 

The mobility in the lateral or in-plane direction, in a layered 
heterostructure, is substantially enhanced through selectively doping 
specific regions [5]. Consider a quantum well made up of two layers of 
compound semiconductor alloys, acting as the barrier regions, on either side 
of a layer with a smaller energy bandgap. On doping the barrier regions with 
impurities, the bound carriers are released into the quantum well which has 
accessible energy levels that are lower than the impurity donor energy 
levels in the barrier region. If the barrier region has impurity atoms only 
on one side, the released free carriers are confined to a narrow region 
close to the barrier-well interface, forming a two-dimensional electron gas 
(2DEG) [6]. The local separation of the charges leads to substantial bending 
of the conduction band-edge, as in figure 1. The carriers occupy sub-band 
energy levels in the quantum well and are held in the layer by the 
positively charged ionized donors left behind in the barrier. The ionized 
impurity scattering in lateral transport in the quantum well layer, due to 
the placing of the impurities in the barrier regions, can be substantially 
decreased further by set-back doping [7].

In this paper, we are concerned with calculating the dependence of the 
mobility on the spacer layer thickness $d_{s}$, quantum well thickness 
$d_{w}$, temperature $T $ and the doping density $N_{d}$ in a delta-doped layer in 
the barrier. The use of delta-doping with a spacer layer leads to a very 
large mobility for the carriers. We also work with the AlInSb-InSb system in 
which the band gap and the electron effective mass in the lateral direction 
are small, which is an advantage for obtaining enhanced mobility as high as 
4.1 m$^{2}$/V$\cdot $s at room temperature [8], 20.9 m$^{2}$/V$\cdot $s at 
77 K [8], and 28.0 m$^{2}$/V$\cdot $s at 7 K [9], in ultra thin-film 
structures. This attractive property based on the presence of the InSb layer 
has been used in obtaining extraordinarily high magnetoresistance (EMR) in 
semiconductor-metal hybrid structures[4]. While a number of papers in the 
literature treat the AlGaAs-GaAs [10-14]$^{ }$system theoretically, there 
are only a few that deal with transport properties of the AlInSb-InSb 
structures [15].
\begin{figure} 
 \centering 
 \includegraphics[width=0.30\textwidth]{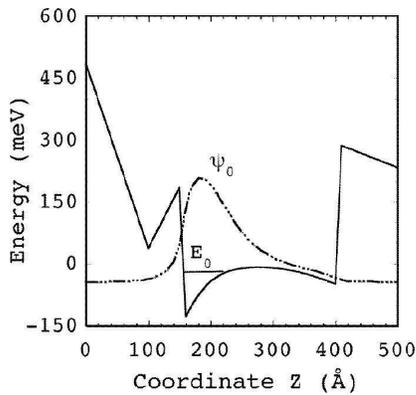} 
 \caption{Figure 1. The conduction band bending, the ground state energy level and its wave function in 
 the $\delta$ -doped layered heterostructure are shown.} 
 \label{...} 
\end{figure}

In the following, in Sec. II, we describe the calculation of the energy 
levels in the heterostructure by solving Schr\"{o}dinger's equation, and the 
corresponding wavefunctions, together with the solution of the Poisson 
equation for the band edge potential given the redistribution of the 
charges. In Sec. III, we evaluate the carrier mobility by determining the 
lifetimes for scattering from the impurities and from phonons. Concluding 
remarks are relegated to Sec. IV.
\section{Solving the Schr\"{o}dinger and Poisson equations}
\begin{figure} 
 \centering 
 \includegraphics[width=0.27\textwidth]{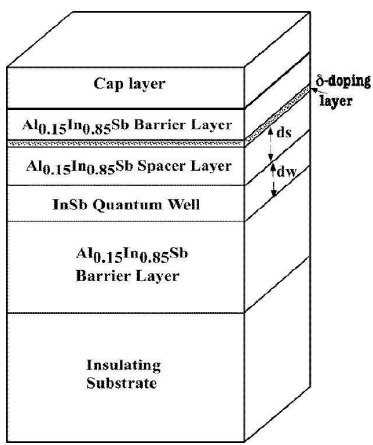} 
 \caption{Figure 2. A schematic diagram of a  $\delta $-doped Al$_{0.85}$In$_{0.15}$Sb-InSb heterostructure. } 
 \label{...} 
\end{figure}
We first provide an outline of the method. The Schr\"{o}dinger and Poisson 
equations are solved iteratively by first starting with the original band 
edge profile with no band bending. The wavefunctions obtained for the 
electron in the original potential are used to determine the charge 
distribution of the electrons, and this carrier charge density together with 
the ionized donor charge density at the delta-doped layer are used as the 
source terms for Poisson's equation. The change in the band edge profile due 
to the redistribution of the charges is solved for and a fraction of the 
change (under-relaxation) in the potential is added to the Schr\"{o}dinger's 
equation to initiate the next iteration. At each iteration, the Fermi level 
is determined through the charge neutrality of the structure as a whole. 
Convergence is reached when the potential function $V$ changes by less than 0.1 
meV over the last iteration [12]. \\

\textit{2.1 Solution of the Schr\"{o}dinger and the Poisson equations}\\

We consider the AlInSb-InSb structure shown in figure 2. The donors in the 
$\delta $-doped layer release electrons into the quantum well layer. These 
electrons are held close to the quantum well interface by the ionized 
impurities left behind and form a 2DEG in the triangle-shaped potential 
created by this redistribution of the charges. 

In the effective mass approximation [16-17] we assume that the envelope 
functions of the electrons are propagating solutions in the in-plane ($x, y)$-direction 
and the $z$-dependent envelope function satisfies the one-band equation 
\begin{equation}
-\frac{\hbar ^2}{2}\frac{d}{dz}\left( {\frac{1}{m_z ^\ast }\frac{d}{dz}\psi 
_n } \right)+\left( {E_{CB} \left( z \right)+V\left( z 
\right)} \right)\psi _n 
=\varepsilon _n \psi _n .\tag{2.1}\label{eq:mynum}
\end{equation}
 Here, m$^{\ast }_{z }$is the effective mass in each layer, $E_{CB} $ is 
the original band-edge profile while the potential energy due to the 
redistribution of the charges is $V(z)$. We neglect nonparabolicity [18] of the 
conduction band energy dispersion. We solve the above equation by the finite 
element method (FEM) [19-20]. 

The one dimensional Poisson equation is 
\begin{equation}
\frac{d}{dz}\left( {\varepsilon \varepsilon _r \left( z 
\right)\frac{dV\left( z \right)}{dz}} \right)=-e\rho _e =-e^2\left[ {n\left( 
z \right)-N_d \left( z \right)} \right],\tag{2.2}\label{eq:mynum}
\end{equation}
 where $\varepsilon _r $ is the relative dielectric constant. $N_d $ is the 
ionized donor concentration,$n(z)$ is the electron density in the Hartree 
approximation, and $\rho _e $ is the charge distribution [12, 20]. 
The electron density distribution is related to the wave function and subband electron occupation by

\begin{equation}
n(z)=\sum\limits_{i=1}^{n_d } {n_i ^{2D}\left| {\psi _i \left( z \right)} 
\right|^2} .\tag{2.3}\label{eq:mynum}
\end{equation}
Here $n_i ^{2D}$ is the electron occupation for each state given by
\[
\begin{array}{l}
 n_i ^{2D}=\frac{m^\ast k_B T}{\pi \hbar ^2}\ln \left( {1+e^{\frac{E_f -E_i 
}{k_B T}}} \right){\begin{array}{*{20}c}
 , \hfill & \hfill \\
\end{array} }_ (T>0); \\ 
 n_i ^{2D}=\frac{m^\ast }{\pi \hbar ^2}\Theta \left( {E_f -E_i } 
\right){\begin{array}{*{20}c}
 , \hfill & \hfill & \hfill & {(T=0)} \hfill\tag{2.4}\label{eq:mynum} \\
\end{array} }. \\ 
 \end{array}
\]
where $E_f $ is the Fermi energy and $E_i $ is the subband energy of the 
bound states in the quantum well, and $\Theta $ is the step-function. The 
self-consistent potential $V(z)$ depends on $\rho _e $ which in turn depends on 
$E_i $ and $\psi _i $ obtained from the solution of the Schr\"{o}dinger 
equation.

We seek bound state solutions to the Schr\"{o}dinger equation with $\psi 
(0)=0$ at $z $= 0 at the top of the heterostructure and $\psi (L)=0 $at the 
bottom. At layer interfaces, the wavefunction continuity and the 
mass-derivative of the wavefunction, $\psi '/m^\ast  $, are assumed to be 
continuous [21].The boundary conditions for the Poisson equation are: $V(z=$0) = $V_{c}$, a 
contact potential, and $V(z=L)$ = 0, at the bottom of the heterostructure. At 
layer interfaces, we require that $V$ and $\varepsilon _r dV/dz$ be continuous. 
In order to solve Eqs. (2.1-2.4) self-consistently, we start with the 
initial choice for the potential function, $V $= 0, and solve the 
Schr\"{o}dinger equation for the bound state energies $\varepsilon _n $ and 
wavefunctions $\psi ^{(0)}$.From this we construct the first approximation to the charge density$\rho _e 
$. Now Poisson's equation is solved to determine the change in the potential 
energy, $V(z)$, with the source terms given by the charge density $-e\left[ 
{n\left( z \right)-N_d \left( z \right)} \right]$. A fraction of this 
potential energy function is added to the potential energy used in the 
Schr\"{o}dinger equation. Letting $k$ be the iteration index, we set 
$V^{k+1}\left( z \right)=\alpha \cdot V^k\left( z \right)+\left( {1-\alpha } 
\right)\cdot V^{k-1}\left( z \right)$, where $\alpha $ is $\sim $0.01. This 
is used, in turn, in the Schr\"{o}dinger equation for the next iteration of 
the wavefunction $\psi _n ^{k+1}$ using 
\begin{equation}
-\frac{\hbar ^2}{2}\frac{d}{dz}\left( {\frac{1}{m_z ^\ast }\frac{d}{dz}\psi 
_n ^{k+1}} \right)+\left( {E_{CB} +V^k} \right)\psi _n ^{k+1}=\varepsilon _n 
^{k+1}\psi _n ^{k+1}.\tag{2.5}\label{eq:mynum}
\end{equation}
These wavefunctions are used in defining $n^{k+1}(z)$ in the source term in 
Poisson's equation
\begin{equation}
\frac{d}{dz}\left( {\varepsilon _0 \varepsilon _r \frac{dV^{k+1}}{dz}} 
\right)=-e\rho _e ^{k+1}=-e^2\left[ {n^{k+1}\left( z \right)-N_d \left( z 
\right)} \right].\tag{2.6}\label{eq:mynum}
\end{equation}
The process is repeated until convergence is reached through the criterion 
that the change in the potential function reaches the tolerance $\lambda $ 
such that
\begin{equation}
\frac{\left| {V^{k+1}\left( z \right)-V^{k}\left( z \right)} 
\right|}{\left| {V^{k}\left( z \right)} \right|}\le \lambda .
\tag{2.7}\label{eq:mynum}
\end{equation}
The tolerance was chosen to be $\lambda \cong 1\times 10^{-3}$. The 
resulting conduction band edge profile is shown in figure 1, together with 
the ground state wavefunction.
The conduction band edge $E_{CB} $ in each layer is determined by a 
parameter, $\gamma $, representing the conduction band offset in the layer 
with respect to the band-gap, $E_g $, in the material and a reference layer. 
We include the temperature dependence of the band gap as given by the 
empirical Varshni equation [22]
\begin{equation}
E_g \left( T \right)=E_g \left( {T=0} \right)-\frac{\alpha T^2}{T+\beta 
}\quad ,\tag{2.8}\label{eq:mynum}
\end{equation}
where $\alpha $ and $\beta $ are the usual Varshni parameters [23].\\

For all the ternary alloys, the composition-dependent bandgap is given by a 
simple linear or quadratic equation as a function of concentration $x$ by 
[24-25]
\begin{equation}
E_g \left( {A_{1-x} B_x } \right)=\left( {1-x} \right)E_g \left( A 
\right)+xE_g \left( B \right)-x\left( {1-x} \right)C.\tag{2.9}\label{eq:mynum}
\end{equation}
Where $C$ is the bowing parameter for the bandgap, and accounts for the 
deviation from a linear interpolation between the bandgaps of the two 
binaries $A$ and $B $[23]. \\

\textit{2.2 The determination of the Fermi level}\\

As the electrons fall into the quantum well the band-edge potential in the 
well region moves upwards to make it less energetically unfavorable for the 
next electron to enter the well region. If band nonparabolicity is ignored, 
the 2DEG density is given by [11] 
\begin{equation}
n_s =\frac{\left( {E_f -E_0 } \right)\,m^\ast }{\pi \,\hbar ^2}.\tag{2.10}\label{eq:mynum}
\end{equation}
If the barrier doping density is large, the location of the Fermi level is 
defined by the energy level of the unionized donor and charge neutrality 
requires that 
\begin{equation}
\sum{n_{s} } =\sum{N_{d }} .\tag{2.11}\label{eq:mynum}
\end{equation}
The Fermi level can be determined by using a root-finding procedure. 
Equation (2.11) is solved for the Fermi level in every iteration, in order 
to speed up the convergence. 
\section{Calculation of the mobility}
The carrier mobility, $\mu =e\tau /m^\ast $, is governed by scattering 
mechanisms that control the scattering time, $\tau $, associated with each 
scattering process in a 2DEG [26]. The scattering rates calculated with each 
scattering mechanism are combined to determine the resultant mobility. The 
dominant scattering mechanisms for bulk III-V compounds are now well 
established [27-29]. In our calculations, we have included these mechanisms 
in the context of 2D scattering. \\

\textit{3.1 The scattering from ionized impurities}\\

At low temperature, scattering is dominated by the ionized donors. In 
a $\delta $-doped structure with a spacer layer, the ionized donors and the 
2D electrons are spatially separated, thereby minimizing the scattering. We 
assume that the electron gas is highly degenerate and that the scattering 
occurs only with electrons near the Fermi level. The transport scattering 
rate for a purely 2D electron gas, neglecting the electronic wavefunction 
normal to the plane of the 2D gas, is given by [16] 
\begin{equation}
\frac{1}{\tau _{tr} }=N_{imp}^{2D} \frac{m^\ast }{2\pi \hbar ^3k_F^3 
}\nonumber\\
\int_0^{2k_F } {\left| {\mathop V\limits^\sim \left( q \right)} \right|} 
^2\frac{q^2dq}{\sqrt {1-\left( {\frac{q}{2k_F }} \right)^2} }\tag{3.1}\label{eq:mynum}
\end{equation}
where $k_F =\sqrt {n_s^{2D} \cdot 2\pi } $ is the Fermi wavevector, 
$n_s^{2D} $ is the sheet density of the 2DEG, $N_{imp}^{2D} $ is the 
impurity doping areal density and $\mathop V\limits^\sim \left( q 
\right)=\int {V\left( r \right)e^{-i\mathop q\limits^\to \cdot \mathop 
r\limits^\to }} d^2\mathop r\limits^\to $is the Fourier transform of the 
scattering potential.

Using Thomas-Fermi screening, the scattering rate for remote impurities 
becomes [16] 
\begin{widetext}
\begin{equation}
\frac{1}{\tau _{tr} ^{imp}}=n_{imp}^{2D} \frac{m^\ast }{2\pi \hbar ^3k_F^3 
}\left( {\frac{e^2}{2\varepsilon _0 \varepsilon _r }} \right)^2 \nonumber\\
\int_{0}^{2k_F 
} \frac{e^{-2q\left| d \right|}}{\left[ {q+q_{TF} G\left( q \right)} 
\right]^2}\left( {\frac{b}{b+q}} \right)^6\frac{q^2dq}{\sqrt {1-\left( 
{\frac{q}{2k_F }} \right)^2} },\tag{3.2}\label{eq:mynum}
\end{equation}
\end{widetext}
with
\[
G(q)=\frac{1}{8}\left[ {2\,\left( {\frac{b}{b+q}} \right)^3+3\,\left( 
{\frac{b}{b+q}} \right)^2+3\,\left( {\frac{b}{b+q}} \right)} \right],
\]
and 
$b=\left( {\frac{33\,m^\ast e^2n{ }_{2D}}{8\,\hbar ^2\varepsilon _0 
\varepsilon _r }} \right)^{1/3}$.
The scattering potential takes account of screening with the Thomas-Fermi 
dielectric function [30]. Here $q_{TF} =\frac{m^\ast e^{2}}{2\pi \hbar 
^{2}\varepsilon \varepsilon _{r}}$ is the screening wavevector in the 
Thomas-Fermi approximation. The contribution to the mobility from ionized 
impurity scattering is essentially independent of temperature [31].
\\

\textit{3.2 The ionized background impurities}\\

The background density of ionized donor impurities is usually very small as 
compared with ionized impurities with thin spacer layer thickness. But when 
the thickness of the spacer layer is very large, the effect of remote 
ionized impurities is reduced and the background impurity scattering 
dominates. The rate for scattering from the low density background 
impurities with a concentration of $N_{bg-imp}^{3D} $ is given by [16] 
\begin{widetext}
\begin{equation}
\frac{1}{\tau _{tr} ^{bg}}=N_{bg-imp}^{3D} \frac{m^\ast }{2\pi \hbar ^3k_F^3 
}\left( {\frac{e^2}{2\varepsilon _0 \varepsilon _r }} \right)^2
\int_0^{2k_F 
} \frac{1}{\left( {q+q_{TF} } \right)^2}\frac{qdq}{\sqrt {1-\left( 
{\frac{q}{2k_F }} \right)^2} }.\tag{3.3}\label{eq:mynum}
\end{equation}
\end{widetext}

\textit{3.3 The scattering from phonons}\\

Phonons dominate scattering at high temperature, typically above 60 K [32]. 
The most important phonon scattering processes in general are: ($i)$ 
deformation potential acoustic phonon scattering, (\textit{ii}) polar optical phonon 
scattering, and (\textit{iii}) piezoelectric scattering [33-34]. However, Basu and Nag 
[35] have shown that for InSb, the piezoelectric scattering does not play an 
important role at intermediate temperatures. We will therefore not consider 
it here. We consider the electrons to be quasi-two dimensional while the 
phonons remain quasi-three-dimensional, in which approximation the 
perturbing potential has a spherical symmetry [16]. For the sake of 
simplicity, we assume that in the ideal case we have at hand bulk phonons. 

The deformation potential acoustic phonon scattering rate is given by [32, 
36] 
\begin{equation}
\frac{1}{\tau _{tr}^{acoustic} }=\frac{3m^{\ast} k_{B }T\,\Xi ^{2}}{2\,\rho 
\,v_s^{2} \hbar ^{3}d}.\tag{3.4}\label{eq:mynum}
\end{equation}
where $\Xi $ is the deformation potential, and $\rho ,\,v_s $ are the mass 
density and the velocity of sound.

The scattering rate by polar optic phonons is given by [32-33] 
\begin{equation}
\frac{1}{\tau _{tr}^{polar} }=\frac{k_0 e^2\pi N_{0} }{2\hbar \varepsilon_{p} 
},\tag{3.5}\label{eq:mynum}
\end{equation}
with $N_0 =\left( {e^{\frac{\hbar \omega _{0} }{k_B T}}-1} \right)^{-1}$ , 
$k_0 =\sqrt {\frac{2m^\ast \omega _{0} }{\hbar } } $, and 
$\frac{1}{\varepsilon _{p }}=\frac{1}{\varepsilon _\infty 
}-\frac{1}{\varepsilon _{r} }$, with $\varepsilon _\infty $ being the high 
frequency dielectric constant and $\varepsilon _{r} $ being the static 
dielectric constant. Here $\hbar \omega _{0}$ is the optical-phonon 
energy. Several values (7.2 eV and 30 eV [37-38]) have been quoted in the 
literature for the deformation potential in InSb. Since the mobility is 
inversely proportional to $\Xi ^{2}$, these different $\Xi $ values can lead 
to a large difference in the acoustic phonon scattering contribution to the 
mobility. We use a value of 7.2 eV for  $\Xi $  in InSb, which 
was first obtained by Ehrenreich [39] and Dutta [38] who showed that it 
agrees well with the experimental results. \\

\textit{3.4 Other scattering mechanisms}\\

The alloy-disorder scattering by free carriers was neglected in this 
calculation. Interface (or surface) roughness scattering is another 
scattering mechanism which has been found to be important in thin quantum 
wells [40], playing a significant role only at high electron concentrations 
[41], and thin quantum wells ( L$<$60 $\AA$) [42]. When the first excited 
subbands become filled with electrons, there will be interface scattering, 
which would reduce the mobility [43].\\

\textit{3.5 The combined mobility }\\

The total mobility is given by computing the mobility for each scattering 
process and adding the reciprocal mobility for each process according to 
Mathieson's rule. 
\begin{equation}
\frac{1}{\mu _{total} }=\frac{1}{\mu _{imp} }+\frac{1}{\mu _{bg} 
}+\frac{1}{\mu _{acoustic} }+\frac{1}{\mu _{polar} }.\tag{3.6}\label{eq:mynum}
\end{equation}
At high temperature, the relation may not be valid due to the limited 
applicability of degenerate statistics [31]. However, for temperatures 
higher than 60 K the scattering in the 2DEG is dominated by the phonon 
scattering processes. The approximation has a negligible effect on the 
combined electron mobility. Therefore we still can use Mathieson's rule to 
roughly estimate the combined mobility.

The mobility contributions caused by ionized impurity and background 
impurity scatterings at low temperatures are temperature independent, above 
the temperature at which the total mobility shows the T$^{-1}$ dependence.
\section{ Results and discussion}
We studied the Al$_{0.}$85In$_{0.15}$Sb-InSb system. Some relevant 
parameters used in the calculation are given in Table 1. The other material 
parameters employed here have been obtained from Ref. 23.
\textbf{Table 1.} Parameters used in the calculation are shown.
\begin{table}[htbp]
\begin{center}
\begin{tabular}{|l|l|}
\hline
Parameters of InSb& 
Value \\
\hline
Effective mass ratio ($m^{\ast }$/m$_{0})$& 
0.0135 $^{a)}$ \\
\hline
Static dielectric constant ($\varepsilon _{0} $ )& 
16.82 $^{a)}$ \\
\hline
High-frequency dielectric constant ($\varepsilon _\infty $ )& 
15.7 $^{b)}$ \\
\hline
Lattice density (g/cm$^{3})$& 
5.79 $^{b)}$ \\
\hline
Velocity of longitudinal elastic waves (cm/s)& 
3.7$\times $10$^{5} \quad ^{b)}$ \\
\hline
Acoustic deformation potential (eV)& 
7.2 $^{c)}$ \\
\hline
Optical phonon energy (meV)& 
25.0 $^{b)}$ \\
\hline
\end{tabular}\\
\label{tab1}
\end{center}
\end{table}\\
$^{a)}$ Ref. 23 
$^{b)}$ Ref. 15
$^{c)}$ Ref. 37\\

\textit{4.1 Temperature dependence of the electron mobility}
\\

\begin{figure} 
 \centering 
 \includegraphics[width=0.26\textwidth]{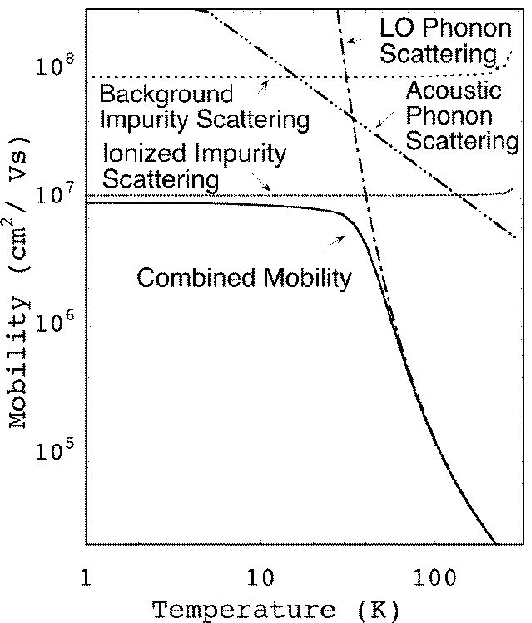} 
  \includegraphics[width=0.31\textwidth]{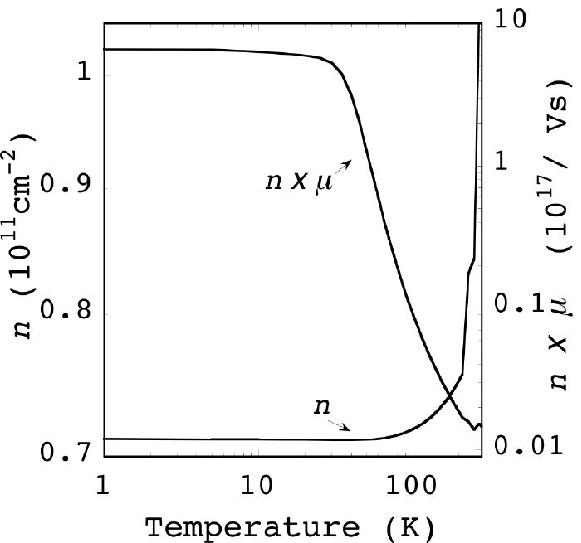} 
 \caption{Figure 3.  The temperature dependence of the electron mobility
  in the Al$_{0.85}$In$_{0.15}$Sb-InSb system is shown. (a) The curves show the calculated 
  mobility with scattering due to the remote ionized impurities with a sheet density of 1$\times$10$^{12}$ cm$^{ Ð2}$. 
  The spacer layer thickness and well thickness are 400 $\AA$.  The background impurity density is
   assumed to be 5$\times$10$^{15}$ cm$^{-3}$. (b) The temperature dependence of the electron subband density
    and the product of the carrier density and the mobility, $n\times $\textit{$\mu $}, are shown.} 
 \label{...} 
\end{figure}
The calculated electron mobility, in the range of 1-280 K, is given for the 
AlInSb-InSb system in figure 3. The components of the mobility contributing 
to the total mobility are also presented. In figure 3(a), with T decreasing 
from 280 K, the carrier mobility increases monotonically and saturates 
around 30 K. This is in contrast to the bulk semiconductor, where the 
mobilities tend to peak at intermediated temperature. This is due to the 
fact that at low temperature, the electron mobility is limited by ionized 
impurity scattering and phonon-limited scattering is negligible; on the 
other hand, at high temperature polar-optical-phonon scattering is the 
dominant scattering mechanism. At intermediate temperature, the 
deformation-potential acoustic phonon scattering also plays an important 
role. At high temperature, the mobility shows the expected $T^{-1}$ 
dependence. The subband electron occupation is shown in figure 3(b). In our 
case, only the ground state subband is filled. The electron density in the 
quantum well remains constant at low temperature. But it increases rapidly 
about 100 K. We know that for the sensor application, high electron density 
determines the conductivity and this is as important as the need for high 
electron mobility. We have therefore plotted the product of electron density 
$n $ and mobility \textit{$\mu {\rm }$} in figure 3(b). The product of $n\times $\textit{$\mu $} decreases rapidly at 
high temperature, and remains constant at low temperature. \\

\textit{4.2 Spacer thickness dependence }\\

\begin{figure} 
 \centering 
 \includegraphics[width=0.27\textwidth]{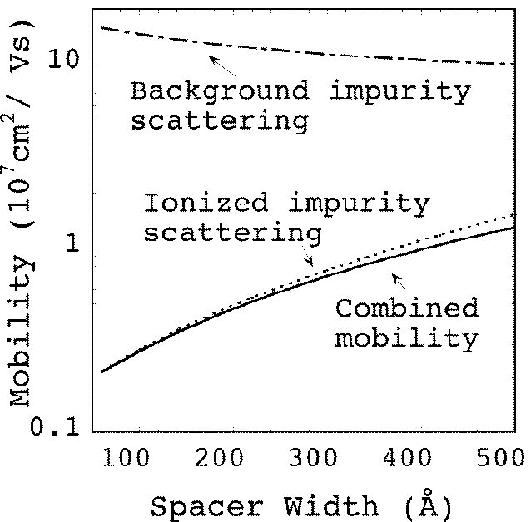} 
  \includegraphics[width=0.31\textwidth]{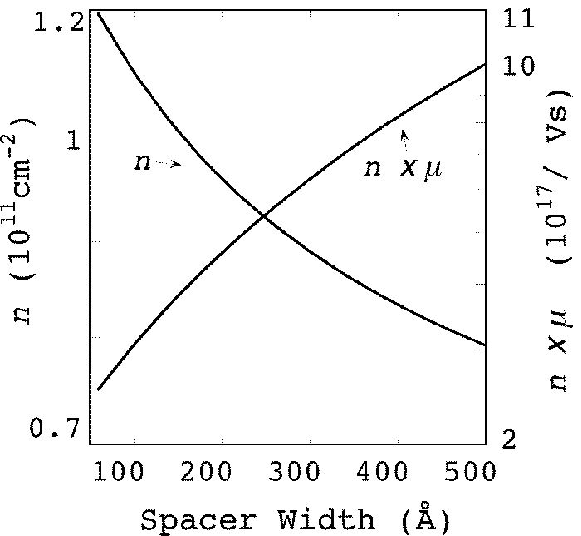} 
 \caption{Figure 4. (a)The dependence of the electron mobility on the thickness of the spacer layer in
  the Al$_{0.85}$In$_{0.15}$Sb-InSb heterostructure is shown. The calculation is performed at T=0 K for a well width
   of 200 $\AA$. The background impurity density is assumed to be 5$\times$10$^{15 }$cm$^{-3}$. (b) The dependence of electron 
   subband density and the product of the carrier density and the mobility, $n\times $\textit{$\mu $}, on the spacer layer thickness are shown.} 
 \label{...} 
\end{figure}
The purpose of a $\delta$-doped heterostructure is to separate 
the 2DEG from the parent ionized donors, thereby limiting ionized impurity 
scattering from the remote doping centers. Figure 4(a) shows that an 
increase in the spacer thickness leads to an increase in the combined 
mobility. The ionized impurity mobility increases with increasing spacer 
thickness, reaching mobilities of 1x10$^{7}$ m$^{2}$/V$\cdot $s at a spacer 
width of 400 {\AA}. The mobility caused by the introduction of background 
ionized impurities shows the reverse trend. The background impurity mobility 
decreases with increasing spacer width and becomes dominant. However, an 
increase of spacer layer thickness leads to a decrease in the electron 
density $n$ in the quantum well as shown in figure 4(b). The product \textit{n$\times \mu {\rm }$} and the 
carrier density $n$ are also shown in figure 4(b), with the former increasing 
with spacer layer thickness.

For a background density of 5$\times $10$^{15}$ cm$^{-3}$, the spacer width 
dependence of combined mobility is insensitive to the background impurity as 
can be seen in figure 4(a). However, higher background impurity 
concentrations will make the background impurity mobility comparable with 
the ionized impurity mobility which will result in the combined mobility 
displaying a peak at some spacer thickness, as shown in figure 5. Figure 
5(b) also shows that \textit{n$\times \mu {\rm }$} as a peak value at a spacer width around 300 {\AA} at 
high background impurity density.\\

\textit{4.3 Dependence on the quantum well width }\\

\begin{figure} 
 \centering 
 \includegraphics[width=0.30\textwidth]{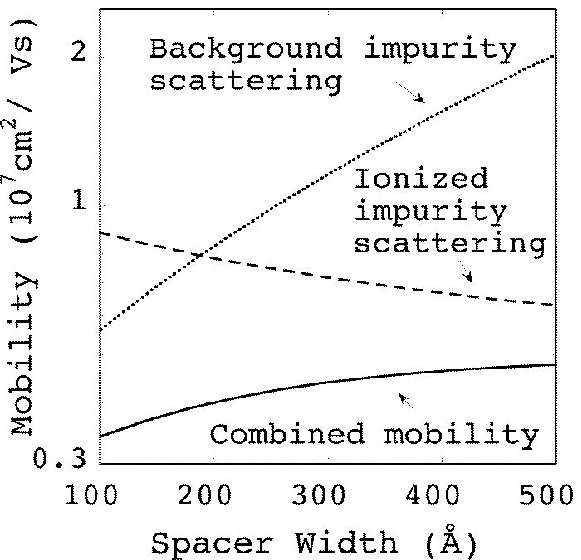} 
  \includegraphics[width=0.30\textwidth]{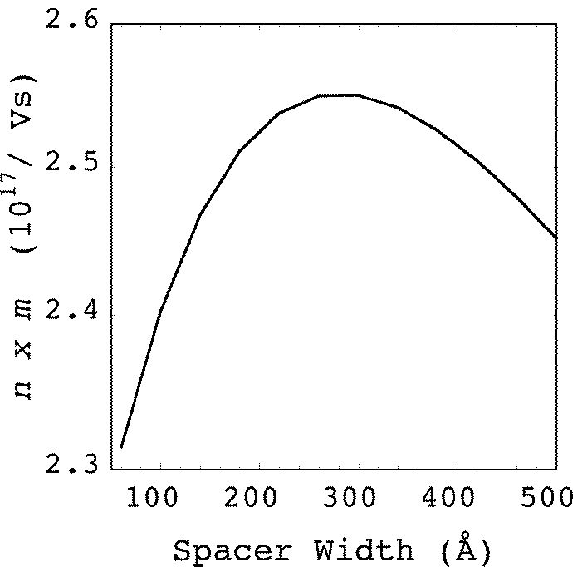} 
 \caption{Figure 5.  (a) The dependence of the electron mobility on the spacer layer thickness in the 
 A$_{l0.85}$In$_{0.15}$Sb-InSb heterostructure is shown. The background impurity density is assumed to be 5$\times$10$^{16}$cm$^{-3}$.  
 (b) The product of the carrier density and the mobility, \textit{n$\times \mu {\rm }$}, is shown as a function of the spacer layer thickness.} 
 \label{...} 
\end{figure}\begin{figure} 
 \centering 
 \includegraphics[width=0.29\textwidth]{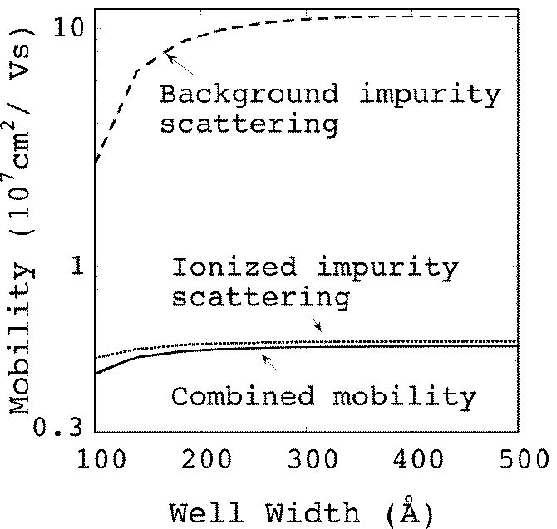} 
  \includegraphics[width=0.30\textwidth]{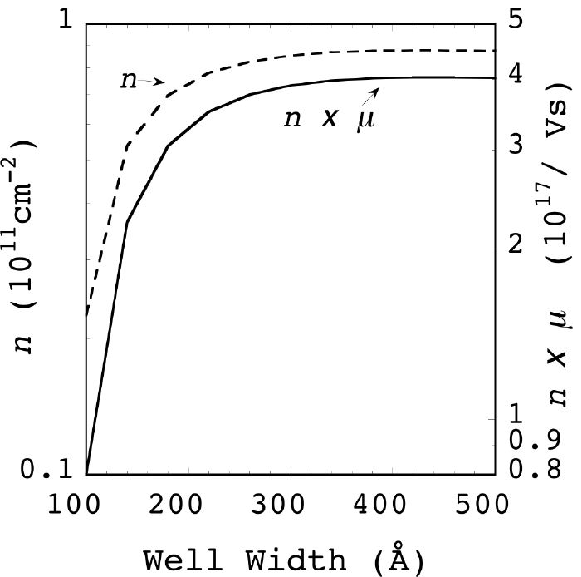} 
 \caption{Figure 6.  (a) The dependence of the electron mobility on the well width in the Al$_{0.85}$In$_{0.15}$Sb-InSb heterostructure 
 is shown. The calculation is performed at T=0 K and a spacer layer thickness of 200 $\AA$. A background impurity density of 
 5$\times$10$^{15}$ cm$^{-3}$ is assumed. (b) The dependence of electron subband density, and of the product of the mobility and electrons, \textit{n$\times \mu {\rm }$}, 
 on the well layer thickness is shown.
} 
 \label{...} 
\end{figure}
\begin{figure} 
 \centering 
 \includegraphics[width=0.30\textwidth]{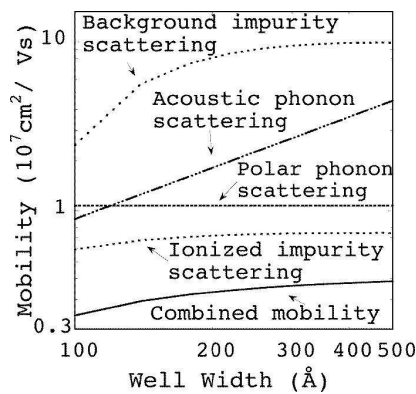} 
 \caption{Figure 7. The dependence of the electron mobility on the well layer thickness in the Al$_{0.85}$In$_{0.15}$Sb-InSb heterostructure is shown on a log-log
  plot, so that the dependence of polar phonon scattering on well width is shown to be nearly constant. The calculation is performed with T=40 K
   and a spacer thickness of 300 $\AA$.
} 
 \label{...} 
\end{figure}

\begin{figure} 
 \centering 
 \includegraphics[width=0.28\textwidth]{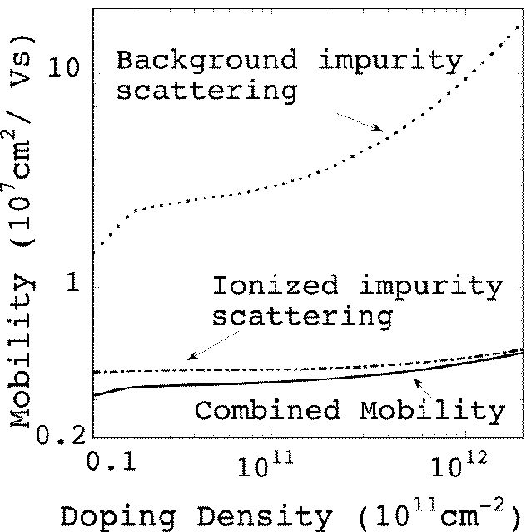} 
  \includegraphics[width=0.32\textwidth]{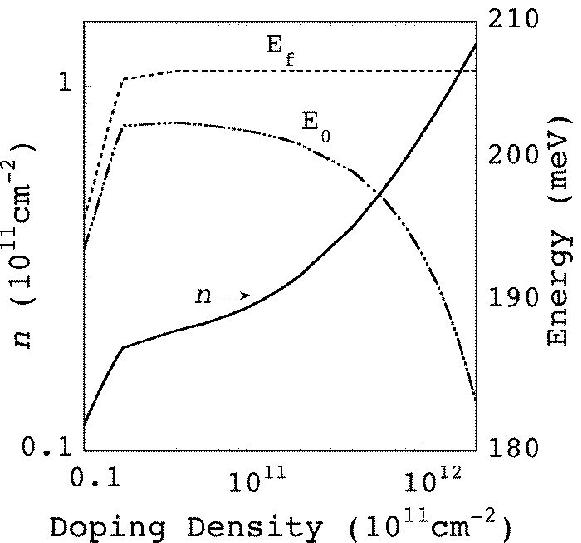} 
    \includegraphics[width=0.28\textwidth]{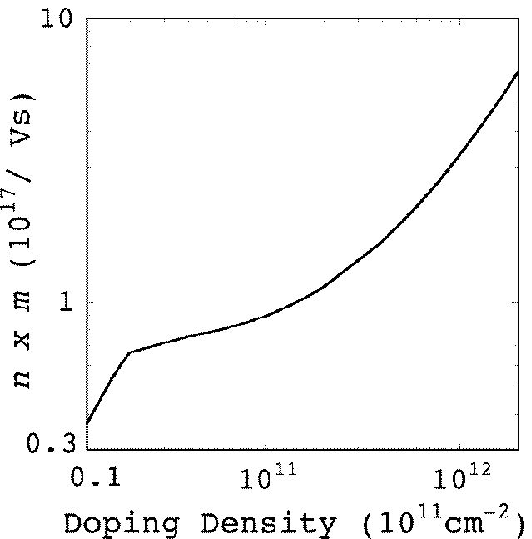} 
 \caption{Figure 8. (a) The dependence of the electron mobility on the doping density in the Al$_{0.85}$In$_{0.15}$Sb-InSb heterostructure is shown. 
 The calculation is performed with T=0 K and spacer and a well thickness of 200 $\AA$. The background impurity density is assumed to 
 be 5$\times$10$^{15 }$cm$^{-3}$. (b) The variation of the Fermi level, the ground subband energy and electron densities with the ionized doping density are
  shown. (c) The product of mobility and electron density, \textit{n$\times \mu {\rm}$}, versus the doping density is shown.
} 
 \label{...} 
\end{figure}
For the well width dependence study we considered the system with a constant 
spacer layer thickness of 200 {\AA}. Figures 6(a) and (b) show that at 0 K 
for a thin quantum well ( $<$ 200 {\AA}), electron density $n $and \textit{n$\times \mu {\rm }$} increase 
rapidly up to $d_{w }$\textit{$\sim $ }200 {\AA}, while at larger well width, they increase 
slowly and saturate at a well width 300 {\AA}. With the increase in the well 
width, the background impurity mobility increases faster than the ionized 
impurity mobility and is a less important factor in the combined mobility. 
The electron density $n $ and the product \textit{n$\times \mu {\rm }$} show the same trend with increasing well 
width. 

Equations (3.4)-(3.6) show that the acoustic-phonon mobility should increase 
linearly with the well width at intermediate temperature, and that the polar 
optic phonon scattering remains constant. Figure 7 shows the calculated 
mobility at 40 K. Background impurity and ionized impurity mobility show a 
similar trend compared to the 0 K mobility (Fig. 6a). \\

\textit{4.4 Doping density dependence}\\

Figure 8 shows that at 0 K as the remote impurity density increases, the 
background impurity mobility increases faster than the ionized impurity 
mobility. This follows logically, because with more electrons in the quantum 
well, the effect of background impurities is not the limiting factor in the 
combined mobility, therefore it shows a trend similar to that of the 
electron subband density in figure 8(b). At low doping density, all the 
electrons can enter the well without pushing the Fermi energy up to the 
conduction band edge. With increased doping density, more electrons can 
enter the well, and the Fermi energy reaches the conduction band-edge and is 
unable to move up further. The ground subband energy will then be lower to 
let more electrons into the well as shown in figure 8(b).
\section{Conclusion}
In conclusion, we have presented theoretical calculations for the electron 
subband energy, subband electron density, and carrier mobility in $\delta$
-doped single quantum well heterostructures. Here, the Hartree 
approximation for the confinement potential was used. The important 
scattering mechanisms, such as, ionized impurity scattering, background 
impurity scattering, deformation potential acoustic phonon scattering, and 
polar phonon scattering were considered. We have presented the behavior of 
the product of electron mobility and density in the temperature range of 0 - 
280 K as a function of the spacer layer thickness, well width, and doping 
density.

At low temperature, the dominant scattering mechanism is ionized impurity 
scattering. It is also found to be independent of temperature in this 
regime. In the AlInSb-InSb system, the mobility due to ionized impurity 
scattering increases with decreasing temperature, reaching a limiting value 
of about 1x10$^{7}$ cm$^{2}$/V$\cdot $s for a thicker well and larger spacer 
( $>$300 {\AA}) at low temperature ( 0 K). At a temperature around 40 K, 
optical phonons begin to limit the mobility which varies as T$^{-1 }$at high 
temperature. The dependence of \textit{n$\times \mu {\rm }$} shows a maximum for a spacer thickness of 300 
{\AA} for higher background impurity densities while it continues to 
increase monotonically for lower background impurity densities.

\begin{acknowledgments}
This work was supported by the NSF (Grant No. ECS-0329347) and the Center 
for Materials Innovation at Washington University. LRR thanks Washington 
University in St. Louis for the Harrison Fellowship which supported his 
collaborative visit during part of which this work was carried out.
\end{acknowledgments}


\end{document}